\documentclass[document]{IEEEtran}

\usepackage{bm}
\usepackage{multicol}
\usepackage{graphicx}
\usepackage{latexsym}
\usepackage{amsmath}
\usepackage{hyperref}
\usepackage{pxfonts}
\usepackage{cancel}
\usepackage{bbding}
\usepackage{subfig}
\usepackage{float}
\usepackage{subfloat}
\usepackage{caption}
\usepackage[section]{placeins}
\DeclareGraphicsExtensions{.pdf,.png,.jpg}

\begin{document}

\small {

\title{\LARGE{Relativistically Induced Transparency Acceleration (RITA) of Light-ions by an Ultra-short Laser Interacting with a Heavy-ion Plasma Density Gradient}}

\author{Aakash A. Sahai*, \thanks{e-mail: (aakash.sahai@gmail.com)}
        Frank S. Tsung,
        Adam R. Tableman,
        Warren B. Mori,
        Thomas C. Katsouleas
\thanks{A. A. Sahai and T. C. Katsouleas are with Department of Electrical Engineering, Duke University, Durham, NC, 27708 USA}
\thanks{F. S. Tsung, A. R. Tableman and W. B. Mori are with Department of Physics and Astronomy, University of California, Los Angeles, CA, 90095 USA}%
}
\vspace{0mm}

\maketitle

\begin{abstract}
The Relativistically Induced Transparency Acceleration (RITA) scheme of proton and ion acceleration using laser-plasma interactions is introduced, modeled and compared to the existing schemes.
Protons are accelerated with femtosecond relativistic pulses to produce quasi-mono-energetic bunches with controllable peak energy. 
The RITA scheme works by a relativistic laser inducing transparency\cite{rit-1970-1971} to densities higher than the cold-electron critical density, while the background heavy-ions are stationary. The rising laser pulse creates a traveling acceleration structure, at the relativistic critical density by ponderomotively\cite{ponderomotive-1999} driving a local electron density inflation, creating a snowplow and a co-propagating electrostatic potential. The snowplow advances with a velocity determined by the rate of the rise of laser's intensity envelope and the heavy-ion plasma density gradient scale length. The rising laser is incrementally rendered transparent to higher densities such that the relativistic-electron plasma frequency is resonant with the laser frequency. In the snowplow frame, trace density protons reflect off the electrostatic potential and get snowplowed while the heavier background-ions are relatively unperturbed. Quasi-mono-energetic bunches of velocity equal to twice the snowplow velocity can be obtained and tuned by controlling the snowplow velocity using laser-plasma parameters. An analytical model for the proton energy as a function of laser intensity, rise-time and plasma density-gradient is developed and compared to 1-D and 2-D PIC OSIRIS\cite{osiris-code-2002} simulations. We model the acceleration of protons to GeV energies with tens of femto-seconds laser pulse of a few PetaWatts. The scaling of proton energy with laser power compares favorably to other mechanisms for ultra-short pulses\cite{tnsa-2006}\cite{rpa-2004}\cite{cesa-2004}.  
\end{abstract}

\begin{figure}
	\begin{center}
   	\includegraphics[width=3.6in]{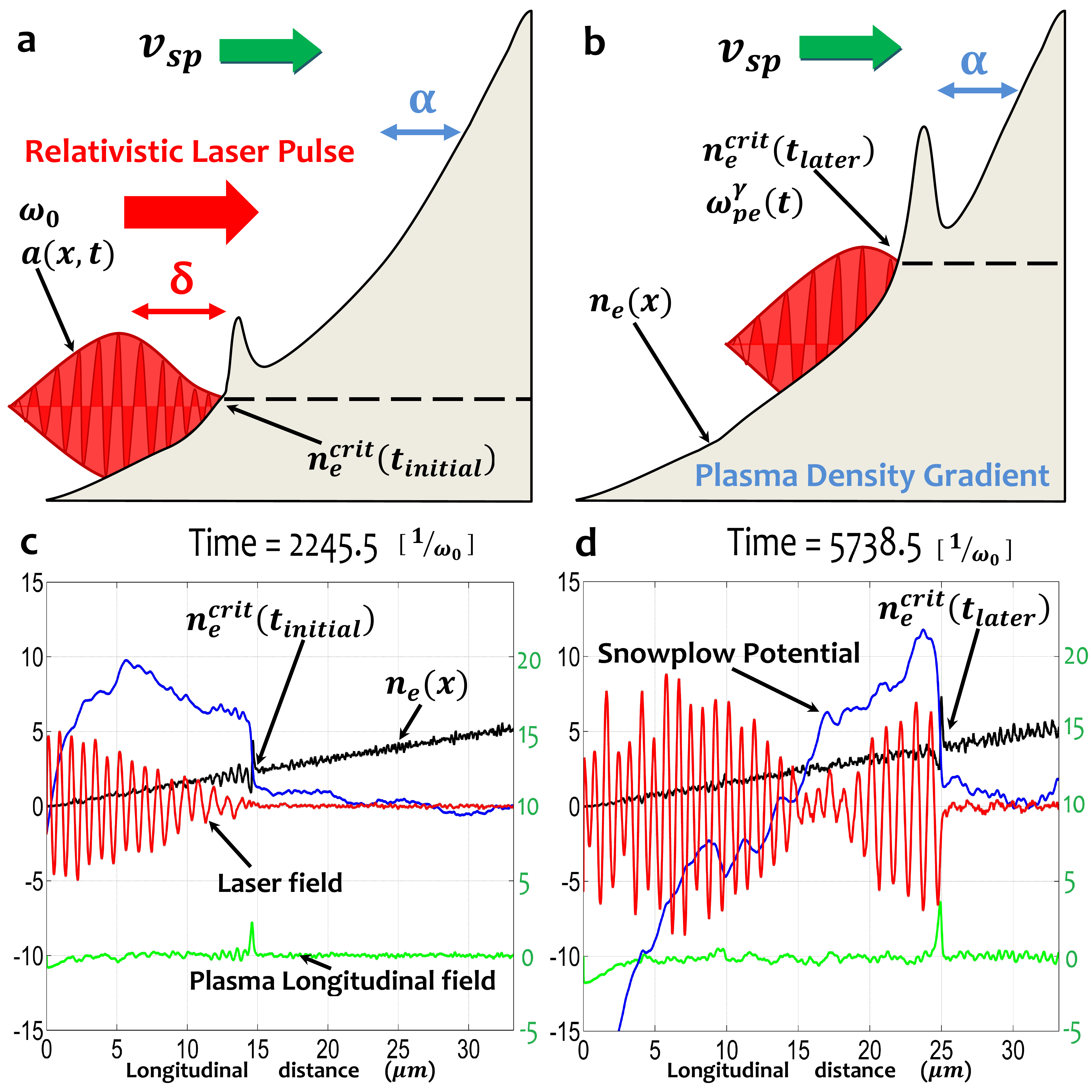}
	\end{center}
\caption{ { \scriptsize {\bf Snowplow formation and its evolution with fixed-ion 1-D PIC simulations}. {\bf a,} Snowplow formation at the critical density,  $n_e^{crit}(t_{initial})$ at $t_{initial}$. The plasma density scale length is $\alpha$; relativistic laser pulse rise-time is $\frac{\delta}{c}$. {\bf b,} The propagation into uphill plasma density to $n_e^{crit}(t_{later})$, with velocity, $v_{sp}$ due to the rising intensity in the head of the laser pulse envelope. {\bf c,} Fixed ion 1-D PIC simulation at $t_{initial}=2245.5\frac{1}{\omega_0}=954fs$ for a continuously rising laser intensity with scale length $\delta=1904\frac{c}{\omega_0}\rightarrow\frac{\delta}{c}\simeq 810fs$, $a_0=2$ and with the plasma density rise scale length $\alpha=50\frac{c}{\omega_0}=6.4\mu m$ ($\lambda_0=0.8\mu m$). The laser electric field is in red, plasma longitudinal electric field is in green (axis on the right) and snowplow potential, $\Phi_{sp}$ is in blue (all in normalized units). {\bf d,} Corresponding, 1-D simulation at a later time $t_{later}=5738.5\frac{1}{\omega_0}=2437fs$. From (c,d), $v_{sp}\simeq0.024c$, as in Fig.\ref{fig:Snowplow-velocity-scaling-laws}. It should be noted these simulations are carried out with immobile background plasma heavy-ions. The entire movie corresponding to the snapshots is in the supplemental material \cite{phase-space-movies}. } }
\label{fig:Snowplow-concept}
\end{figure}

Laser-plasma accelerators can generate proton and ion beams with unprecedented characteristics of high bunch charge, low emittance and ultra-short bunch lengths using a very compact system. However, existing mechanisms for acceleration \cite{tnsa-2006}\cite{rpa-2004}\cite{cesa-2004} with femto-second lasers have been unable to simultaneously achieve the desired energy gain and spectral control needed for the applications of interest.  Here we demonstrate with analysis and PIC simulations \cite{osiris-code-2002} that using heavy-ion targets, we can control the laser propagation by means of relativistically induced transparency \cite{rit-1970-1971}. By varying the laser rise-time or plasma density gradient it is possible to tune the energy of the mono-energetic beams. The results provide a pathway for achieving the beam properties needed for a wide-array of applications ranging from hadron therapy \cite{proton-therapy-2002} to high-charge injectors \cite{compact-injectors-2000}, particle physics \cite{fission-plasma-ions-2003} and high energy density physics \cite{fast-ignition-fusion-2001}.


Experimentally, target normal sheath acceleration (TNSA) scaling laws have been studied extensively\cite{scaling-laws-nphys}\cite{scaling-laws-2010-ultrashort}, and low-energy quasi-mono-energetic beams demonstrated\cite{tnsa-mono-2006}. The radiation pressure (RPA)\cite{rpa-mono-solid}\cite{rpa-mono-co2} and collisionless electrostatic shock acceleration (CESA)\cite{cesa-mono-2012} have been experimentally demonstrated with low energy quasi-mono-energetic bunches using picosecond TeraWatt $CO_2$ lasers or nanometer-scale targets. This motivates the RITA mechanism which exploits the use of relativistically-intense femtosecond lasers interacting with heavy-ion targets for high-efficiency proton and light-ion acceleration. This letter is organized into an introduction of the scheme including the development of scaling laws. The analytical model is then compared to PIC OSIRIS \cite{osiris-code-2002} simulations. Lastly, we compare the RITA scaling to TNSA.

The RITA scheme is depicted in Fig.\ref{fig:Snowplow-concept}[a][b]. An intense laser pulse propagates up an appropriately graded heavy-ion plasma. When the head of the laser pulse with electric field oscillating at an angular frequency $\omega_0\leq\omega_{pe}^{\gamma\simeq 1}=\sqrt{\frac{4\pi e^2n_e}{m_e^{\gamma\simeq 1}}}$, eq.(1) ($n_e$ is plasma electron density) reaches the cold-plasma critical density $n^{crit}_{cold}=\frac{m_e^{\gamma\simeq 1}\omega_0^2}{4\pi e^2}$, it is reflected back. As the laser intensity rises, the quiver transverse motion of the electrons in the laser field becomes relativistic (kinetic energy of the electrons in the laser field becomes greater than their rest-mass energy), increasing their mass by the Lorentz factor, $\gamma^{\perp}_{e}=\sqrt{1+\frac{\vec{p}_{\perp}.\vec{p}_{\perp}}{m_e^2c^2}}=\sqrt{1+|\vec{a}|^2}$, eq.(2) thereby modifying the cold-electron plasma frequency, $\omega_{pe}^{\gamma}(x,t)=\sqrt{\frac{4\pi e^2n_e(x)}{\gamma^{\perp}_{e}(x,t)~m_e^{\gamma=1}}}=\sqrt{\frac{4\pi e^2n_e(x)}{\left(\sqrt{1+|\vec{a}(x,t)|^2}\right)~m_e^{\gamma=1}}}$, eq.(3). The relativistic-electron plasma density is $\frac{n_e(x)}{\gamma^{\perp}_{e}(x,t)}$. Where, $\vec{a}(x,t)=\frac{e\vec{A}(x,t)}{m_ec^2}=\frac{\vec{p}^{\perp}_e(x,t)}{m_ec}=\gamma^{\perp}_e(x,t)~\vec{\beta}^{\perp}_e(x,t)$, ($|\vec{a}_{peak}|=a_0$) is the normalized vector potential and $\vec{A}(x,t)$ is the laser field vector potential in vacuum. Thus relativistic intensity $|\vec{a}|^2 > 1$, allows the laser to propagate further into the uphill plasma density gradient. This process of reduction of relativistic-electron plasma frequency ($\omega_{pe}^{\gamma}$) compared to cold-electron plasma frequency ($\omega_{pe}^{\gamma\simeq 1}$) for the same plasma density with fixed background-ions ($M_{ion}$) is known as relativistically induced transparency \cite{rit-1970-1971}. It should be noted for clarity that ultra-short laser pulse intensity (with pulse length, $\tau_p \ll \sqrt{m_p/m_e}/\omega_0 \ll \sqrt{M_{ion}/m_e}/\omega_0$, where $m_p$ is the mass of proton) varies in space-time and has a characteristic envelope with a rising part, a flat-top (not always) and a falling part. Since, the laser can propagate further into the plasma density gradient by inducing transparency only if it can render higher plasma densities transparent, an advancing front can occur only while the laser intensity is rising. An observed consequence of this effect is the optical shuttering\cite{rit-expt-2012} (experimental connotation of relativistic transparency) of a laser pulse, which when more intense than a threshold intensity, propagates beyond an initially reflecting ultra-thin nanometer-scale boundary, more by density reduction due to expansion of plasma\cite{rit-expt-expanding-plasma-1992} than due to the relativistic effects.

In addition to the transverse dynamics of the plasma electrons in the laser field, the changing laser intensity in the envelope results in longitudinal dynamics. The ponderomotive force\cite{ponderomotive-1999}, $\vec{F}_{x}^{~ponde}\simeq -\frac{m_ec^2}{2\gamma}\vec{\nabla}_x{a^2(x,t)}$, eq.(4) of the rising laser pushes away the critical layer electrons, creating a dense local build-up of the electron density just beyond the critical layer\cite{rit-threshold-2000}, referred to as a snowplow (seen in the black curve corresponding to electron density in Fig.\ref{fig:Snowplow-concept}[c][d]). The transparency is incrementally induced such that the relativistic critical layer (and snowplow) is at a density, $n^{crit}_{\gamma}(x,t)$ where the relativistic-electron plasma frequency is equal to the laser frequency $\omega_{pe}^{\gamma}(x,t)\simeq\omega_0$ and $n_e(x)=\gamma^{\perp}_{e}(x,t) ~ n^{crit}_{cold}$, eq.(5). Thereby, the laser transfers energy resonantly to the relativistic critical density ($n^{crit}_{\gamma}(x,t)$) plasma electrons. The process of laser being able to incrementally render higher densities (initially reflecting) transparent to itself continues until $a(x,t)\le a_{max}$ ($a_{max}$ can be higher than free-space peak amplitude of the laser field, $a_0$ due to various plasma effects). The rate at which the transparency is induced is higher if the laser intensity rises faster. The velocity of the snowplow $v_{sp}$ therefore depends on the rate of rise of the laser pulse intensity and the plasma density gradient scale length. The electron snowplow forms only under the condition that its speed is less than the group velocity of the laser ($\frac{v_g}{c}=\sqrt{1-\frac{n_e}{n^{crit}_{cold}\gamma_e^{\perp} }}$) , and is able to keep up with the incrementally advancing transparency condition above. For laser pulses with rise-time too short or plasma with density gradient scale-length too long, the electron-snowplow is not created and the laser reaches its transparency limit without incrementally inducing the transparency or interacting resonantly with increasing density. For short pulse lasers, the massive background-ions ($M_{ion}$) are relatively unperturbed by the laser ponderomotive force or the electrostatic pull of the inflated electron layer. As a result their inertia establishes the local plasma electron density that controls the rate of laser advance. The spatial charge separation between heavy-ions and electrons in the region depleted of electrons just before the snowplow, sets up a propagating electrostatic potential $\Phi_{sp}$ that follows the snowplow. If the potential difference is large enough, then a positively charged third species of protons or light ions of trace density (at least an order of magnitude below cold critical density, $n_{cold}^{crit}$), of mass $m_p$ can be picked up and accelerated to speeds twice that of the snowplow (reflected off the electrostatic potential in the moving frame of reference of the snowplow) gaining a kinetic energy of $\frac{1}{2}m_p[2\times v_{sp}]^2$, Fig.\ref{fig:Snowplow-proton-reflection-concept}[a]. By appropriately controlling the laser rise-time and the plasma density gradient, tunable beams of quasi-mono-energetic protons or light ions can be produced. If the third species of ions to be accelerated is cold enough initially such that its thermal kinetic energy is well below the electrostatic potential energy of the snowplow, all of it can be picked up and driven longitudinally (hence the analogy to a snowplow). The transverse electric field of the laser in the plasma ($a_{plasma}(x,t)$) can be maximally coupled to a propagating space-charge electrostatic potential ($\Phi_{sp}$) only if maximum charge separation can be created in the snowplow density inflation of electrons against the stationary heavy-ions. Experimentally, the pre-plasma gradient can be controlled through the properties of laser pre-pulse\cite{preplasma}.

The RITA mechanism differs from previous laser-plasma ion accelerator schemes in several very important aspects. The heavy background-ions are essentially immobile over the RITA timescales, whereas the ion motion is critical to both hole boring\cite{absorption-1992} and shock acceleration\cite{cesa-2004}. RITA thereby differs from hole-boring in that no hole is created in the heavy-ion species and from shock acceleration since an ion acoustic shock cannot be launched. Both the hole-boring and shock speed depend upon the plasma-ion mass, so under infinite mass limit these processes are ruled out. Whereas the snowplow velocity $v_{sp}$ in RITA (where ions are stationary) is not dependent upon mass ratio of the background-ions and electrons, $\frac{M_{ion}}{m_e}$. The experimental demonstration of hole-boring\cite{rpa-mono-co2} and shock\cite{cesa-mono-2012} acceleration have been performed in the ion-motion regime of picosecond pulse-length lasers using electron-proton plasma and not with heavy-ion metal plasma and femtosecond lasers as considered here. RITA also differs from TNSA which occurs at the vacuum-plasma interface on the rear of the target at later timescales when the front surface fast electrons reach it and the plasma slowly expands into vacuum. In mobile-ion simulations we see all mechanisms present. When $v_{sp}>v_{shock}$, which occurs when plasma density gradient scale length is sufficiently long, only snowplow is seen. Later, we observe the TNSA field form due to the ponderomotively driven electrons on the front side reaching the plasma-vacuum boundary on the high-density side (seen in Fig.\ref{fig:2D-TNSA-shock}). This TNSA field then marginally further accelerates the RITA accelerated protons when they reach the plasma-vacuum boundary. An electron density movie of these processes at different time-scales is in \cite{phase-space-movies}. At much later times after the laser pulse has reached its peak and is effectively stopped from propagating further ($n^{crit}_{max(\gamma)}$) into the plasma, we can see the formation of a freely propagating shock from the background-ion phase-space (seen in Fig.\ref{fig:2D-TNSA-shock}).

\begin{figure}
	\begin{center}
   	\includegraphics[width=3.6in]{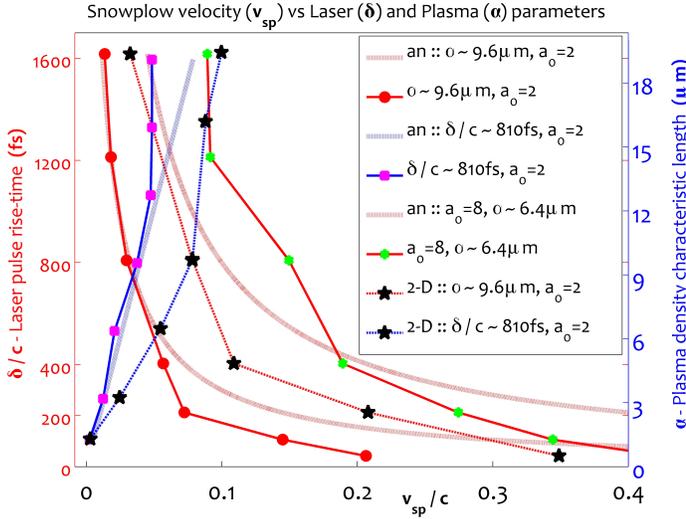}
	\end{center}
\caption{ {\scriptsize {\bf 1-D simulation results of $v_{sp}$ vs. laser-plasma control parameters.} Snowplow velocity $v_{sp}$ vs. laser pulse rise-time $\frac{\delta}{c}$ for various fixed $\alpha$ (in red) and vs. plasma density scale-length $\alpha$ for various fixed laser pulse rise-times $\frac{\delta}{c}$ (in blue). The curves labelled 'an` correspond to, $v_{sp}\simeq 0.5 \left(\frac{\alpha}{\delta}\right)a_0~c$. Also, shown are some corresponding 2-D $v_{sp}$ results for $a_0=2$. Again, it should be noted that these simulations are with immobile plasma background ions.} }
\label{fig:Snowplow-velocity-scaling-laws}
\end{figure} 

To understand the scaling laws, we obtain an analytical expression for the snowplow velocity by equating equation (3) to the laser angular frequency, $\omega_{pe}^2(x,t)=\frac{\omega_{pe}^2(x)}{\gamma_e^{\perp}(x,t)}=\omega_0^2$, eq.(6) and varying it with respect to $x$ and $t$ (laser could be chirped\cite{chita-snowplow} so $\omega_0$ also depends on $x$ and $t$). This variation gives, $\frac{\delta x}{\delta t}=\frac{1}{2}\frac{\partial a^2(x,t)}{\partial t} \left(\frac{1+a^2(x,t)}{\omega^2_{pe}(x)}\frac{\partial \omega^2_{pe}(x)}{\partial x} -\frac{1}{2}\frac{\partial a^2(x,t)}{\partial x}\right)^{-1}$, eq.(7) using the resonance between local plasma frequency and the laser frequency. To simplify, we assume the uphill plasma density to be linearly rising with scale length $\alpha$, as $n_e(x) = n_{cold}^{crit}\left(\frac{x}{\alpha}\right)$, eq.(8) (when $x=\alpha$, the density corresponds to the cold-plasma non-relativistic critical surface); and the laser pulse intensity to be linearly rising, with the rise-time scale length $\delta$ (rise-time = $\frac{\delta}{c}$), as $a^2(x,t) = a_0^2\left(\frac{ct-x}{\delta}\right)H(ct-x)$, eq.(9) (where H is the Heaviside step function). Upon substituting the linear models in eq.(8) and eq.(9) into eq.(7), we get $v_{sp}(t)=\frac{\alpha}{\delta}\frac{a_0^2}{2}\left( \frac{\alpha}{\delta}\frac{a_0^2}{2} + \sqrt{1+a^2(x,t)}\right)^{-1}c$, eq.(10). 
For $a^2(x,t)\gg 1$ and $\frac{\alpha a_0^2}{2 \delta}\ll |a(x,t)|$, $v_{sp}(t)\simeq\frac{\alpha a_0^2}{2\delta}\frac{1}{a(x,t)}c\simeq 0.5 \left(\frac{\alpha}{\delta}\right)a_0c$, eq.(11). In this limit, it is easier to observe that the snowplow velocity scales directly with plasma density scale length ($\alpha$) and inversely with laser pulse rise-time ($\frac{\delta}{c}$), favoring shorter pulses. This also implies, $v_{sp} \propto a_0$. The simplifying assumptions for the scaling law study of the dynamics in 1-D neglect many real effects (observed in simulations). The simple linear analytical model of the laser intensity in equation (9), neglects the fact that laser intensity reaching the snowplow ($\vec{a}_{plasma}\ne \vec{a}_{vacuum}$) is modified as a result of interference (especially under normal incidence) from the doppler-shifted reflected light \cite{einstein-doppler}\cite{rpa-2004} and Airy swelling \cite{wlkruer-lpi-2003} of the laser field due to the changing group velocity of the light in the density gradient (laser field is seen in the red curve in Fig.\ref{fig:Snowplow-concept}[c][d]). The simple analytical model of the plasma density gradient in equation (8) neglects the ponderomotively driven electron density inflation. The laser encounters relativistic-electron critical density $n^{crit}_{\gamma}(x,t)$ within the density inflation of electron snowplow $n_e^{sp}(x,t)$, before it propagates up to the unperturbed relativistic-electron critical density as assumed by the linear density gradient model. So, the laser propagates only until it encounters the condition $n^{crit}_{\gamma}(x,t)=n_e^{sp}(x,t)$, not where the $n^{crit}_{\gamma}(x,t)=n_e(x)$. As the ponderomotive force at the propagating relativistic critical layer increases the local electron snowplow density becomes significantly higher than the initial electron density ($n_e^{sp}(x,t) > n_e(x)$). Thereby the locally inflated electron density reduces the relativistic transparency of the laser and hence the snowplow speed. It is important to note that the time-scale of motion of snowplow is much shorter than any significant motion of the background heavy-ions and the ion density is unperturbed, so the electron density inflation creates a large local charge imbalance. This effect is similar to the striction non-linearity when ions are dragged along with the electrons due to radiation pressure observed with flat-top pulses and homogeneous plasma density\cite{rit-threshold-2000}. Additionally the 1-D process in the equation above does not account for the increase in the $\vec{a}(x,t)$ due to relativistic self-focussing of the laser pulse in a reducing skin-depth ($c/\omega_{pe}$) plasma, which focusses the incident laser power into smaller spot-sizes within plasma filaments of size $\sqrt{\gamma_e^{\perp}}\frac{c}{\omega_p}$\cite{filamentation-Kaw-1973}. 

We model the snowplow formation, propagation and its subsequent acceleration of the trace density ion species using the multi-dimensional PIC OSIRIS code \cite{osiris-code-2002}, shown in 1-D simulation snapshots in Fig.\ref{fig:Snowplow-concept}[c][d]. These simulations were done with fixed plasma background ions to model the heavy-ions which do not move over RITA time-scales. In Fig.\ref{fig:Snowplow-concept}[c] the electron snowplow (the local inflation in the black curve) is at a point where the laser (laser field is shown in red) initially stops around $2.5n_{cold}^{crit}$ (around $15 \mu m$, $n_{cold}^{crit}$ is at x = $6.4\mu m$). In Fig.\ref{fig:Snowplow-concept}[d] at a later time and higher $a(x,t)$ the rising laser has induced transparency to its earlier reflecting density and along with the snowplow has advanced to a higher density $n_{\gamma}^{crit}(x,t)$. The simulation is setup with laser-plasma analytical models ($\frac{\delta}{c}\simeq 810fs$, $a_0=2$, $\alpha\simeq 6.4\mu m$) and the background-ions are immobile. To convert from simulation units to real units we take $\lambda_0=0.8\mu m$ for comparison to Ti:Sapphire laser based experiments. However, for collision-less plasmas all simulations can be scaled keeping the same value of $\frac{\omega_{pe}}{\omega_0}$. In 1-D simulations we use $40\pi$ cells per laser wavelength and 60 particles per species per cell. The electrons, protons and background-ions (when simulated with finite mass) have an initial thermal distribution with average temperature of 5keV. We use open boundary conditions for fields and particles and third-order particle shapes with current smoothing and compensation. 

We run a set of 1-D simulations to examine the dependence of the snowplow velocity $v_{sp}$ on laser rise-time scale length $\delta$, plasma density gradient scale-length $\alpha$ and the peak free-space laser vector potential $a_0$. The simulations are started and the laser-plasma electron evolution is let to evolve in time, while the plasma background-ions are stationary. The snowplow position is recorded at various times and average velocity is determined by $\frac{x_{sp}(t_{later})-x_{sp}(t_{initial})}{(t_{later}-t_{initial})}$. These results are summarized in Fig.\ref{fig:Snowplow-velocity-scaling-laws}. The 1-D simulations reasonably follow the analytical model scaling laws (of eq.11, overlaid on the simulation data curves and labelled as 'an' in Fig.\ref{fig:Snowplow-velocity-scaling-laws}), showing that the snowplow moves faster for shorter rise-time pulses and longer scale length plasma density gradient. We have also shown 2-D simulations for the exact parameters of 1-D simulations except for a realistic laser focal spot-size, $r_0=30\frac{c}{\omega_0}=3.8\mu m$ (2-D effects are discussed in more details below). We have also run these simulations with mobile background ions of mass, $M_{ion}=10m_p$ and found that the snowplow speeds are the same.


\begin{figure*}[ht]
	\begin{center}   	\includegraphics[width=6in]{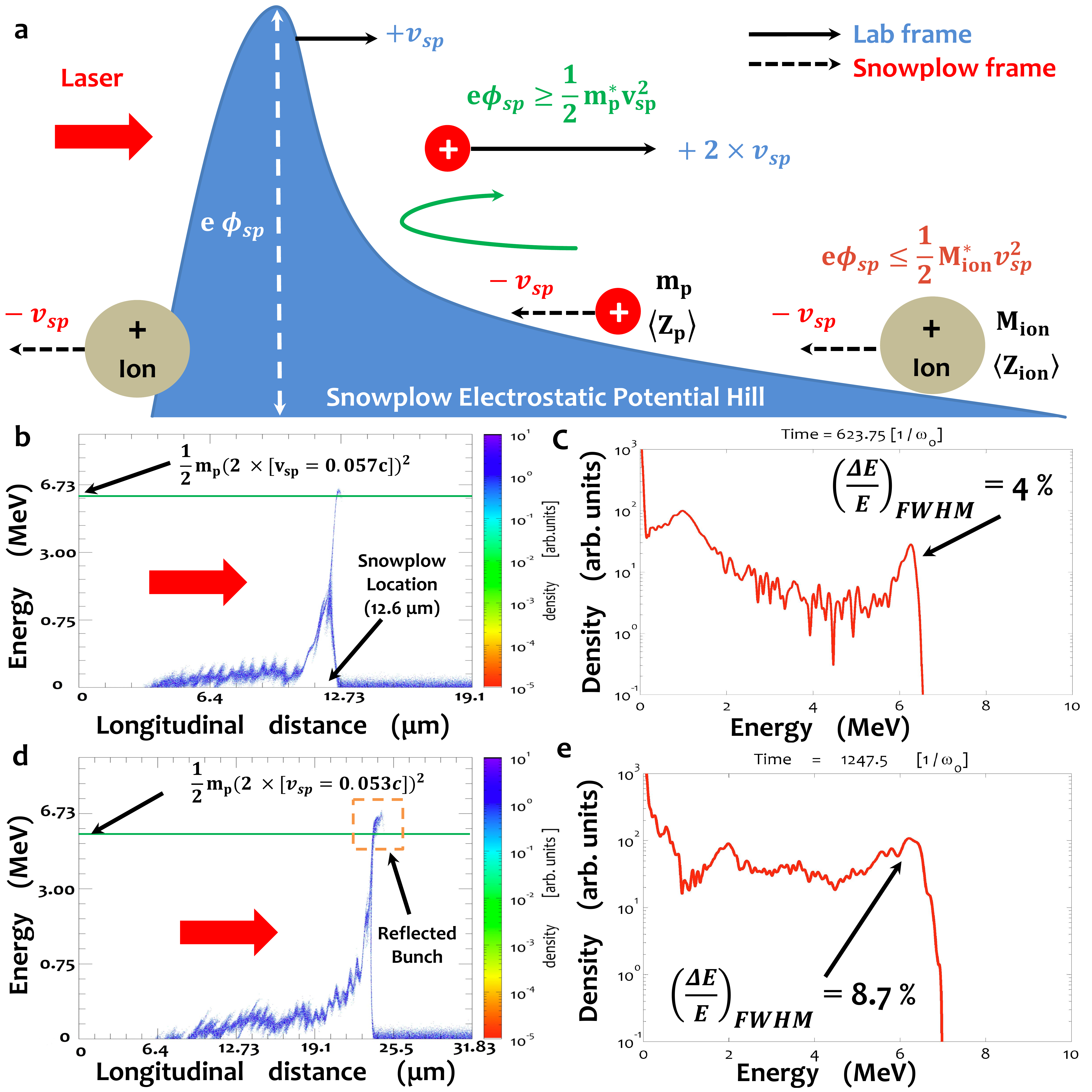}
	\end{center}
\caption{ {\scriptsize {\bf Schematic of acceleration process and proton energy gain from 1-D simulations for different laser-plasma conditions where $\frac{\alpha}{\delta}$ is kept constant.} {\bf a,} Proton reflection off the electrostatic field; protons ($m_p$) gain speed of $2\times v_{sp}$ while background-ions ($M_{ion}$) remain stationary. {\bf b,c,} Longitudinal phase space and energy spectrum of 1-D simulations, (phase space y-axes are in energy units) with $\alpha=6.4\mu m$, $\frac{\delta}{c}=210fs$, $a_0=2$, at $t=623.75\frac{1}{\omega_0}=264.5fs$. It should be noted that in all the energy spectra plots the density is in logarithmic units to assess spectral quality of the proton bunch. The green line is proton predicted energy, $\frac{1}{2}m_p(2\times [v_{sp}=0.057c])^2$. The Full-Width at Half-Maximum (FWHM) energy spread of the bunch is 4\%. {\bf d,e,} $\alpha$ and $\delta$ are doubled ($\alpha=12.7\mu m$ and $\frac{\delta}{c}=404fs$) from (b,c) and shown at $t=1247.5\frac{1}{\omega_0}=529.8fs$. The green line is predicted energy $\frac{1}{2}m_p(2\times [v_{sp}=0.053c])^2$. This shows that similar proton bunch energy gains are obtained by keeping constant $\frac{\alpha}{\delta}$.}}
\label{fig:Snowplow-proton-reflection-concept}
\end{figure*}

Proton and light-ion acceleration occurs when the electrostatic potential reflects them in the snowplow frame, while leaving the heavier background-ions unperturbed (this is shown as a schematic in Fig.\ref{fig:Snowplow-proton-reflection-concept}[a]). The proton reflection starts only when, in the snowplow frame of reference, the protons (or light-ions) with mass-to-charge ratio $m_p^*=\frac{m_p}{\langle Z_p \rangle}$ ($\langle Z\rangle$ is the average ionization state) have a kinetic energy smaller than the electrostatic potential energy of the snowplow hill, $e\Phi_{sp}>\frac{1}{2}m_p^*v_{sp}^2$. The background-ions with mass-to-charge ratio $M_{ion}^*=\frac{M_{ion}}{\langle Z_{ion}\rangle}$ do not get significantly perturbed since $e\Phi_{sp}\ll\frac{1}{2}M_{ion}^*v_{sp}^2$. The reflected protons in the laboratory frame move longitudinally at $2\times v_{sp}$ with a kinetic energy of $\frac{1}{2}m_p(2\times v_{sp})^2$. Thereby the energy gained by the protons does not depend upon the magnitude of the electrostatic potential. However, there is a threshold snowplow potential at which protons start getting reflected.  In the snowplow frame, the threshold laser intensity required to create a large enough snowplow space-charge potential can be estimated by equating the laser ponderomotive potential on the snowplow electrons creating the electrostatic potential $\Phi_{sp}$, to the proton (or light ion) kinetic energy $\frac{1}{2}m_p^*v_{sp}^2$. To estimate the threshold intensity required to develop a large enough charge separation field to allow proton reflection, we consider the ponderomotive potential of a typical electron driven into the snowplow, creating a typical charge separation potential. The snowplow electron energy equation is, $e\Phi_{sp}=(\gamma_{sp} -1) m_ec^2$ where $\gamma_{sp}=\sqrt{1+\left(\frac{|\vec{p}_{\perp}|}{m_ec}\right)^2+\left(\frac{|\vec{p}_{\parallel}|}{m_ec}\right)^2}$ \cite{rit-threshold-2000}\cite{plasma-acc-review-2009}, and $\frac{|\vec{p}_{\perp}|}{m_ec} \simeq \frac{|\vec{p}_{\parallel}|}{m_ec}\simeq |\vec{a}_{plasma}|$ \cite{fast-electrons-pre-2011}\cite{absorption-1992}\cite{absorption-2011}. It should be noted that since $\Phi_{sp}(x,t) \propto a_{plasma}(x,t)$, for a rising laser envelope the traveling electrostatic snowplow potential is increasing. For estimating threshold intensity we equate $(\gamma_{sp}-1)m_ec^2=\frac{1}{2}m_p^*v_{sp}^2$ which under the assumption $|a_{plasma}(x,t)|^2\gg 1$ gives $a_{plasma} > a_{th}\simeq\frac{1}{2\sqrt{2}}\frac{m_p^*}{m_e} \left(\frac{v_{sp}}{c}\right)^2$, eq. (12). Again, determining the local $a_{plasma}(x,t)$ based on the incident $a_{vacuum}(x,t)$ and the scaling of $|\vec{p}_{\parallel}|$ with $|\vec{a}(x,t)|$ is complicated.

The impulse $F_{sp}\Delta t$ of the moving electrostatic potential acting upon the light-ions or protons reflects them and changes their momentum, $\Delta p$. Since the magnitude of the force, $F_{sp}=-e\nabla\Phi_{sp}$ is limited due to a finite slope of the potential, $\Phi_{sp}$ at the snowplow, there is a finite time $\Delta t$ over which the accelerated ions or protons gain the final momentum. Hence, the reflection process is analogous to balls rolling up a hill while losing energy. If their initial kinetic energy is less than the potential of the hill, they roll back down the slope, gaining energy. If their kinetic energy is higher than the potential of the hill they do not reflect off the potential. We observe in our simulations that there is a finite time over which the reflected protons gain their final energy.

There is another important effect associated with the proton and light-ion reflection off the snowplow electrostatic potential. As mentioned above, the laser electric field incident upon the relativistic critical layer is modulated by the Doppler shifted light reflected off the moving snowplow (under normal incidence) forming a beat pattern in the laser pulse envelope as seen in the 1-D simulation snapshot in Fig.\ref{fig:Snowplow-concept}[d]. Due to this beat pattern of the laser envelope the snowplow velocity is modulated, since $v_{sp}\propto a(x,t)$ (from eq. 11). The modulation of the snowplow velocity ($v_{sp}(t)$) leads to the modulation of the reflected ($2\times v_{sp}(t)$) proton spectra (not shown).


We next demonstrate the proton reflection in the 1-D simulations. From simulations we confirm that the proton reflection occurs in the front of the electron snowplow. In Fig.\ref{fig:Snowplow-proton-reflection-concept}[b], proton longitudinal phase space is shown at $264.5fs$, when the proton beam is first formed (with laser-plasma parameters of $\alpha=50\frac{c}{\omega_p}=6.4\mu m$, $\delta\simeq210 fs$ and $a_0=2$). It can be seen that the reflected proton bunch is in front of the electron snowplow location. At this time the vector potential expected from the linear model $a_{vacuum}[264.5fs-\frac{12.62\mu m}{c}]=2.1$ and the threshold is $a_{th}=1.8$ from eq.(12). At the time of proton reflection, from simulations (not shown) the $a_{plasma}(x,t)\simeq 2.4$ and $\Phi_{sp}\simeq 5$. The proton bunch is launched at close to $2\times [v_{sp}=0.057c]$ with a $4\%$ FWHM energy spread as shown in Fig.\ref{fig:Snowplow-proton-reflection-concept}[c].  To demonstrate the proton energy scaling with $\delta$ and $\alpha$, we double $\delta$ and $\alpha$. In Fig.\ref{fig:Snowplow-proton-reflection-concept}[d] a snapshot is shown at $529.8fs$ where model predicted $a_{vacuum}[529.8fs-\frac{23.76\mu m}{c}]=2.23$ and predicted $a_{th}=2.1$; from simulations $a_{plasma}(x,t)\simeq 2.0$ and $\Phi_{sp}=2.92$. As predicted the proton bunch is still approximately at $2\times [v_{sp}=0.053c]$ and has an FWHM energy spread of $8.7\%$, Fig.\ref{fig:Snowplow-proton-reflection-concept}[e]. Thus doubling $\alpha$ and $\delta$ maintains the $\frac{\alpha}{\delta}$ ratio resulting in the same snowplow speed and leaving the proton bunch energy similar under two very different plasma and laser pulse conditions. In these simulations, protons are introduced at a trace density of  $0.01n^{crit}_{cold}$ to model the particle species to be accelerated (the third species) while background-ions are fixed.


\begin{figure}[ht]
	\begin{center}
   \includegraphics[width=3.6in]{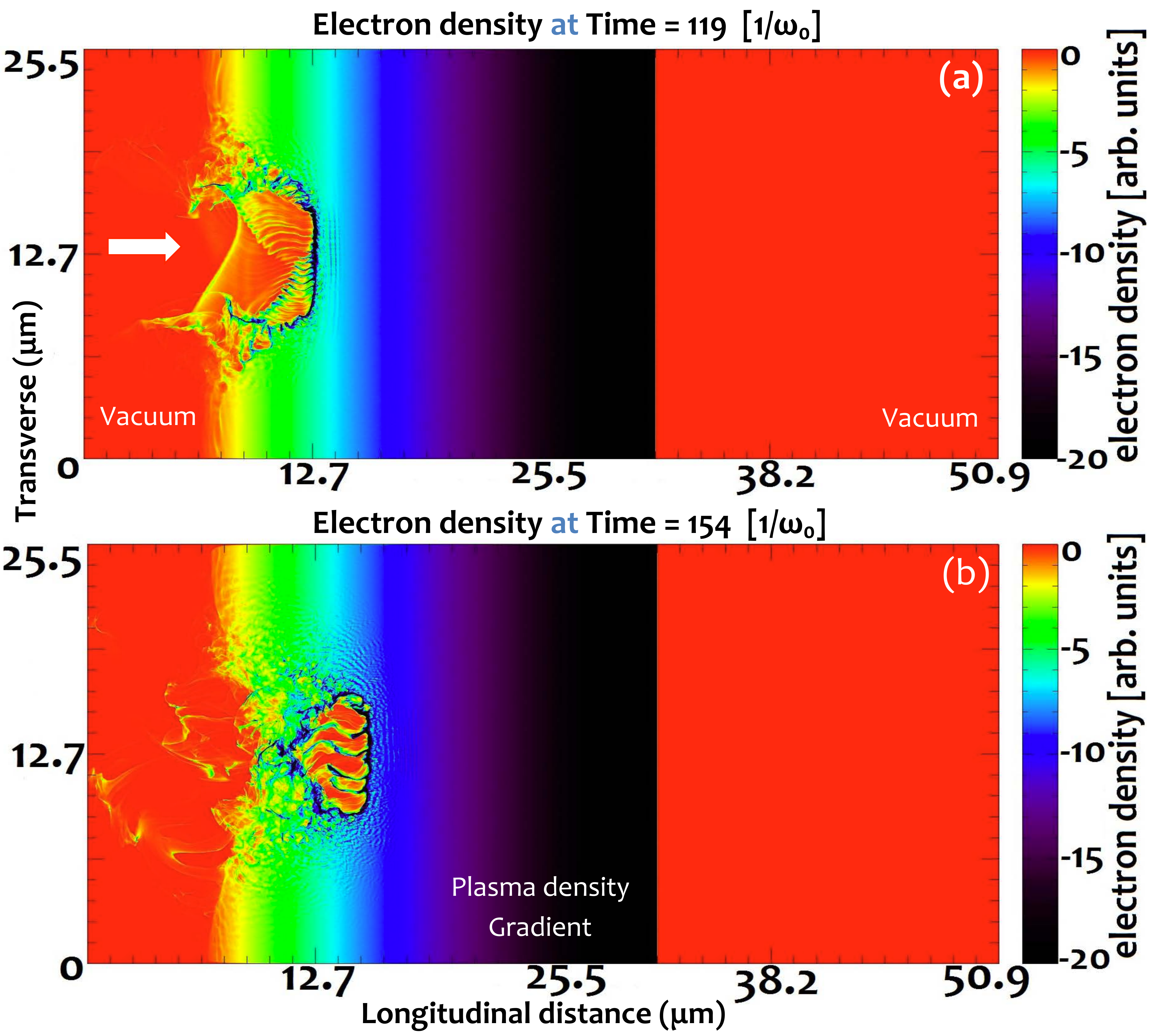}
	\end{center}
\caption{{\scriptsize {\bf Electron density showing the snowplow in 2-D and its evolution.}  {\bf a,} electron density in real space at $119 \frac{1}{\omega_0}\simeq 51fs$, we can observe that an electron density inflation is ponderomotively created by the laser and is located at $\simeq 13.4\mu m$. {\bf b,} snowplow has moved to $\simeq 15.3\mu m$ at $154 \frac{1}{\omega_0}\simeq 66fs$. The plasma density gradient scale length is $\alpha=8.8\frac{c}{\omega_0}=1.12\mu m$. The peak laser intensity is $a_0=72\rightarrow I_0=1.1\times 10^{22}\frac{W}{cm^2}$ and peak power is $2.54PW$. The laser $a(x,t)$ has a Gaussian transverse spatial profile, $a(\vec{r})=a_0~exp(\frac{-r^2}{r_0^2})\hat{r}$ with FWHM focal spot-radius $r_0=30\frac{c}{\omega_0}=3.8\mu m$ and a trapezoidal temporal profile with rise and fall time of $\frac{\delta}{c}\simeq 4fs$ and duration $\tau_p=25fs$. The background-ions have a mass-to-charge ratio, $M^*_{ion}=10m_p$ (high-Z metal like Au, $M_{ion}\simeq 200$ with $\langle Z_{ion}\rangle\simeq 20$ \cite{ionization-state-2005}). The entire movie corresponding to the snapshots is in the supplemental material \cite{phase-space-movies}. }}
\label{fig:2D-snowplow-electron-density-evolution}
\end{figure}

\begin{figure}[ht]
	\begin{center}
   \includegraphics[width=3.6in]{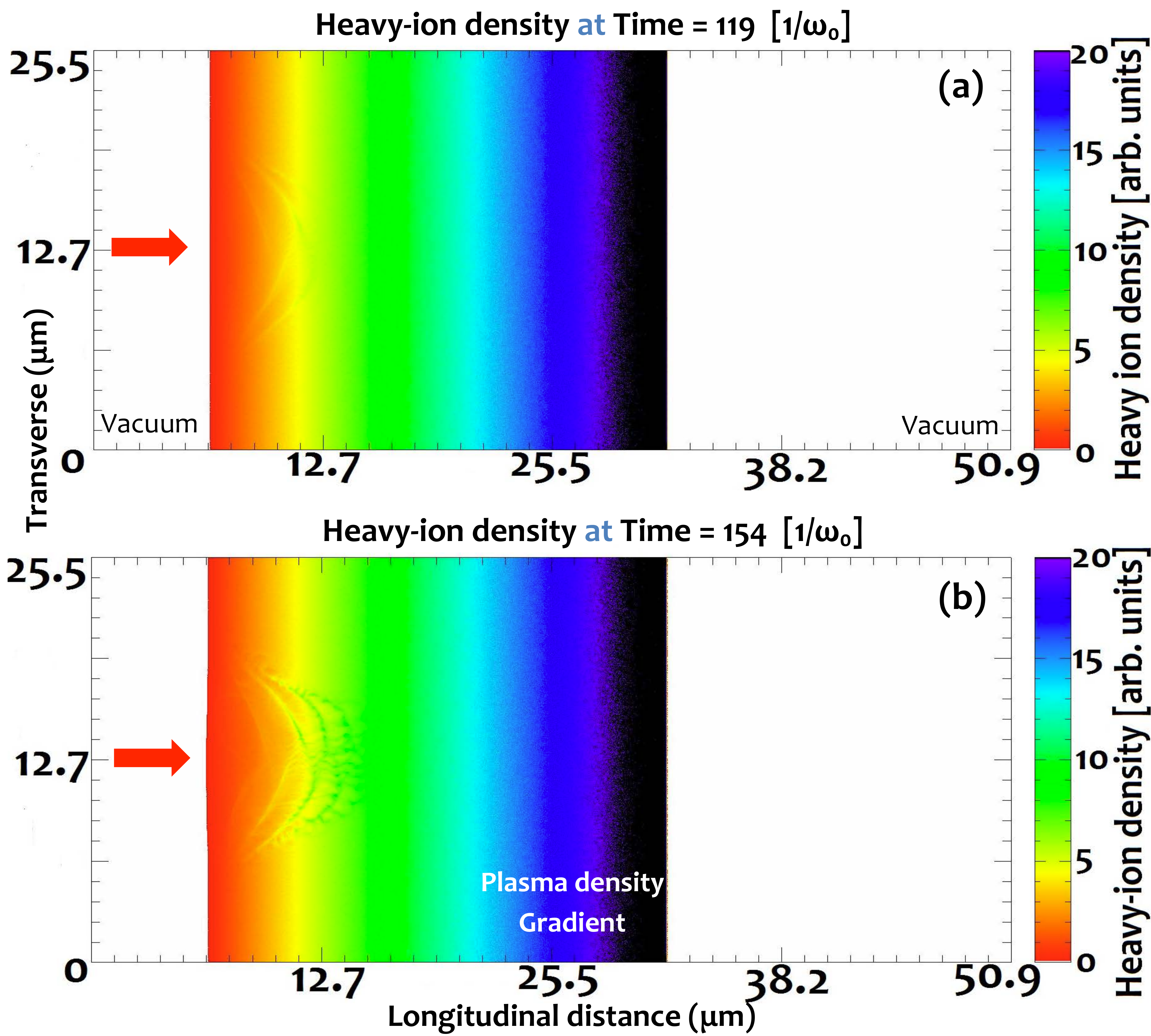}
	\end{center}
\caption{{\scriptsize {\bf Heavy-ion density corresponding to electron snowplow evolution in Fig.\ref{fig:2D-snowplow-electron-density-evolution}.} The evolution of density of background heavy-ions that have a mass-to-charge ratio, $M^*_{ion}=10m_p$ (high-Z metal like Au, $M_{ion}\simeq 200$ with $\langle Z_{ion}\rangle\simeq 20$ \cite{ionization-state-2005}). In particular no perturbation (that is no hole) is seen at the location of the proton acceleration {\bf a,} heavy-ion density in real space at $119 \frac{1}{\omega_0}\simeq 51fs$. {\bf b,} heavy-ion density in real space at $154 \frac{1}{\omega_0}\simeq 66fs$. It should be noted that the laser intensity modeled here is $a_0=72\rightarrow I_0=1.1\times 10^{22}\frac{W}{cm^2}$, which is favorable for heavy-ion motion. But, we still see negligible perturbation. Most of our simulations of RITA are carried out with immobile ions.}}
\label{fig:2D-snowplow-ion-density-evolution}
\end{figure}


We model the signature of RITA scheme with various realizable laser focal spot-sizes using 2-$\frac{1}{2}$ D OSIRIS simulations. We find that with plane waves (equivalent of a 1-D description) and very large focal-spots, the 1-D description of the scheme is complicated in 2-D because of the transverse filamentation of the snowplow\cite{filamentation-Kaw-1973}. Each plasma filament traps a different laser intensity depending on its transverse location with respect to the laser focal spot intensity variation, as show in Fig.\ref{fig:2D-snowplow-electron-density-evolution}. With smaller focal spots of a few laser wavelengths only 2-3 proton filaments form. The relativistic self-focusing of the laser power into filaments with decreasing transverse skin-depth (plasma density is increasing hence the skin-depth $\frac{c}{\omega_p(x)}$ is decreasing), locally increases $a(x,t)$ and the snowplow potential $\Phi_{sp}$ as compared to the 1-D case. Within each filament, the scaling of $v_{sp}$ with $\frac{\alpha}{\delta}$ and of $a_{th}$ follows the analytical theory. But, the snowplow speed and relativistic penetration into higher density\cite{rit-threshold-2000}\cite{rit-timescale-2012} are both less than in simplified model due to the electron snowplow density inflation and other effects in plasma described above. The 1-D simulations corresponding to the 2-D simulations have similar speed but lower penetration. Again, note that the heavy-ion motion (for high intensity case of $a_0=72$) is negligible as seen in Fig.\ref{fig:2D-snowplow-ion-density-evolution}.

\begin{figure*}
	\begin{center}
   \includegraphics[width=6in]{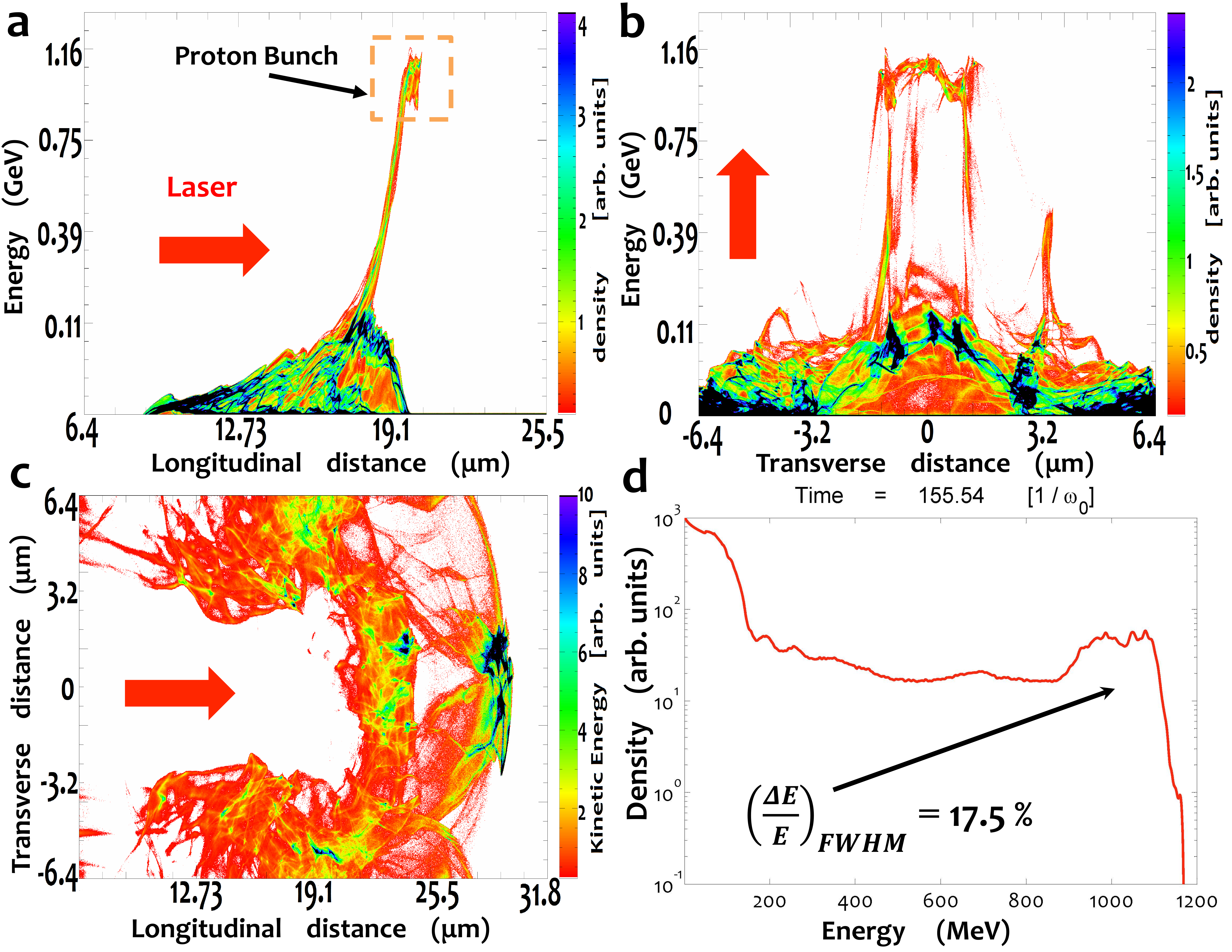}
	\end{center}
\caption{{\scriptsize {\bf Accelerated proton phase-spaces from 2-D simulations in the RITA scheme.} {\bf a,} proton longitudinal phase-space  at $t=220.5\frac{1}{\omega_0}=94fs$ (phase space y-axes are in energy units). From simulations, the time at which the proton bunch is reflected is at $154 \frac{1}{\omega_0}\simeq 66fs$, but the reflected bunch takes finite time $\Delta t$ to reach its maximum energy. {\bf b,} proton longitudinal phase-space in transverse real space. {\bf c,} kinetic energy density in real space at $t=304.5\frac{1}{\omega_0}=129fs$; Note: the appearance of a hole in the proton density should not be confused with hole-boring, since this space is filled with immobile background-ions. A later time is chosen compared to (a) and (b) to show the RITA bunch spatially separated from lower energy protons. {\bf d,} energy spectrum (Note: logarithmic units of the density) with peak at 1.07GeV and FWHM spread of $17.5\%$ at $t=66fs$. The entire movies of different phase spaces corresponding to the snapshots is in the supplemental material \cite{phase-space-movies}.} }
\label{fig:2D-proton-RITA-Snowplow}
\end{figure*}

We show the properties of a quasi-mono-energetic relativistic ($E_p > 938.3 MeV$) proton bunch accelerated by the RITA scheme in Fig.\ref{fig:2D-proton-RITA-Snowplow} using 2-D simulations. It shows the interaction of a circularly polarized laser pulse with peak vector potential of $a_0=72$ (peak intensity is $I_0=1.1\times 10^{22}\frac{W}{cm^2}$ and peak pulse power is $2.54 PW$) of a super-Gaussian transverse spatial intensity profile of FWHM focal spot radius, $r_0=3.8\mu m$ with rising plasma density gradient of scale-length $\alpha=8.8\frac{c}{\omega_0}=1.12\mu m$. It can be seen from Fig.\ref{fig:2D-proton-RITA-Snowplow}[a][b] that the RITA proton bunch is concentrated around 1.07GeV. It can be seen from Fig.\ref{fig:2D-proton-RITA-Snowplow}[c], which plots the kinetic energy density in the real-space, that the length of the ultra-short bunch is $2.5\mu m$ and is of the order of the rise-time of the laser pulse. Since the shape of the laser pulse envelope changes in the plasma due to self-focussing and interfering with the light reflected from the snowplow, the flat-top part of the laser is required to sustain the snowplow electrostatic potential for long enough to allow protons to reflect off the potential hill. It can also be seen from the Fig.\ref{fig:2D-proton-RITA-Snowplow}[c] that the proton bunch is not filamented. The laser temporal profile is a linear rise and fall of $4fs$, with a $17fs$ flat-top part. From 2-D simulations we observe that for laser pulses with sufficiently long rise-times such that the total pulse length $\tau_p\ge 60fs$, the flat-top part is not required. The energy-spectrum of the accelerated protons is shown with the density on a logarithmic scale in Fig.\ref{fig:2D-proton-RITA-Snowplow}[d], which shows an FWHM energy spread of $17.5\%$ at the peak energy $E_p$ of 1.07 GeV.  In this non-optimized 2-D simulation case where proton doping was light $0.01\times n^{crit}_{cold}$, the conversion efficiency of proton energy within the energy spectrum FWHM ($\simeq30\%$ of the total energy in protons) to laser pulse energy is $\simeq 1\%$, corresponding to $\simeq 10^{10}$ protons for a round beam. In 2-D simulations with heavier doping with trace proton density of $0.1\times n^{crit}_{cold}$ shown in Fig.\ref{fig:RITA-point1nc-energy-spectrum}[a], the conversion efficiency reaches as high as $10\%$. Due to heavier beam-loading of the snowplow electrostatic potential in this case $E_p$ is relatively reduced to 0.95 GeV but FWHM energy spread is $10.7\%$. For clarity it should be noted that when the trace species density is set to $\ge n^{crit}_{cold}$, we lose the quasi-mono-energetic signature of the RITA mechanism. We have modeled laser temporal profiles that are Gaussian (not shown) rather than linear and the proton energy spectrum is similar. 

The 2-D PIC simulations are setup with $20\pi$ grid cells per laser wavelength. There are 36 particles per grid cell per species, with 6 particles in each dimension. 
The transverse spatial profile of $\vec{a}$ is Gaussian, $a(\vec{r})=a_0~exp(\frac{-r^2}{r_0^2})\hat{r}$. The background-ions have a mass-to-charge ratio of $\frac{M_{ion}}{\langle Z_{ion}\rangle}=10m_p$. The parameter $a_0$ can be converted to real laser parameters using $I=|\vec{a}|^2\left(\frac{\pi c}{2}\right)\left(\frac{m_ec}{e\lambda}\right)^2$, ($I_{peak}=I_0$), where $r_0$ is the focal spot-size radius and its peak power $P_0=\pi r_0^2\frac{I_0}{2}$.

\begin{figure}
	\begin{center}
   \includegraphics[width=3.6in]{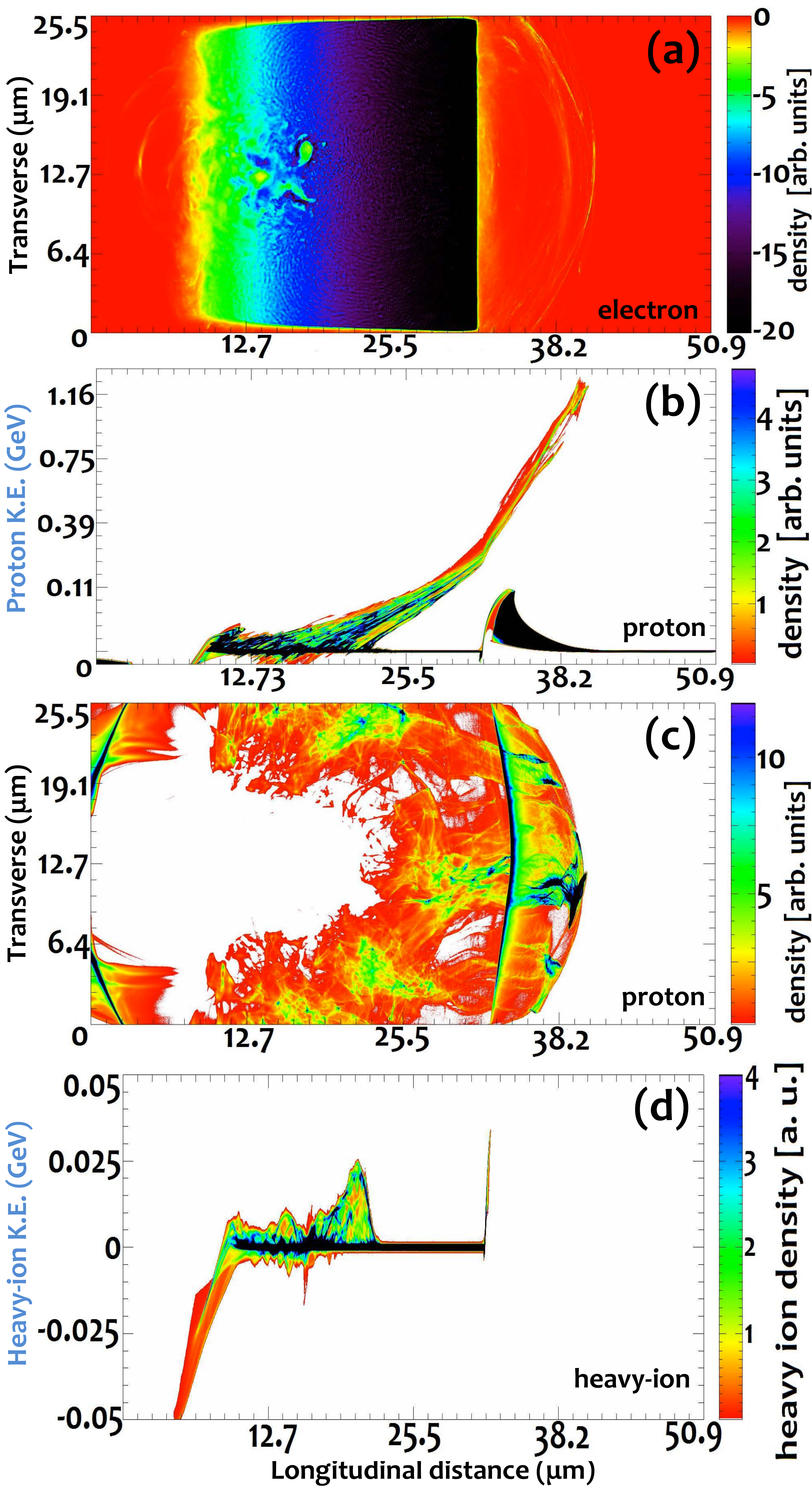}
	\end{center}
\caption{{\scriptsize {\bf Long term evolution of the RITA accelerated protons and the development of TNSA field and shock.} {\bf a,} electron density at $t=385\frac{1}{\omega_0}=164fs$ for laser-plasma parameters in Fig.\ref{fig:2D-snowplow-electron-density-evolution}. At the vacuum-plasma interface at $31.8\mu m$, we can see from the electron density that the electrons expand out away into the vacuum. It should be noted that even though the laser focal-spot FWHM is $7.6\mu m$, the TNSA field transverse size is much larger due to the ponderomotively driven electrons reaching the plasma-vacuum boundary at oblique angles. Also, one can see the shock-like structure in electron density where the snowplow stops around $21\mu m$. {\bf b,} RITA accelerated protons getting a second acceleration from the TNSA fields and gaining higher energy. When compared to the proton longitudinal phase-space in Fig.\ref{fig:2D-proton-RITA-Snowplow}[a] we can observe that the RITA protons have higher energy. Additionally, TNSA accelerated protons can be seen close to the vacuum-plasma boundary, with peak energy of 110 MeV. {\bf c,} Kinetic energy density of the high-energy RITA accelerated protons colliding with the low-energy TNSA accelerated protons. The RITA protons have propagated beyond the TNSA protons and are smaller in transverse dimension. {\bf d,} Heavy-ion phase space showing the TNSA-field at $31.8\mu m $ accelerating the heavy-ions to about 25MeV. Also, the shock field accelerated background heavy-ions can be observed at around $21\mu m$, where the laser and the snowplow finally stop. }}
\label{fig:2D-TNSA-shock}
\end{figure}

To study the effect of the TNSA field excited at the vacuum-plasma interface on the RITA accelerated protons we setup simulations with a steep boundary beyond which the plasma density goes to zero, as seen in plasma density plots in Fig.\ref{fig:2D-snowplow-electron-density-evolution}, Fig.\ref{fig:2D-snowplow-ion-density-evolution} and Fig.\ref{fig:2D-TNSA-shock}[a]. The simulation snapshots in Fig.\ref{fig:2D-TNSA-shock} show the TNSA field phase-spaces and density plots for the simulation parameters corresponding to Fig.\ref{fig:2D-snowplow-electron-density-evolution} and Fig.\ref{fig:2D-snowplow-ion-density-evolution}. The snowplow eventually stops when the snowplow electron density reaches the relativistic critical density. In the simulations the snowplow stops around $22\mu m$. When the snowplow stops there is a dense build up of ponderomotively driven electrons that constitute the snowplow which thermalize and exchange their energy with heavy background ions. Thereby heavy-ions pick up a small momentum (with a maximum kinetic energy of 25MeV) as seen in Fig.\ref{fig:2D-TNSA-shock}[d]. When the laser stops there is also a significant population of ponderomotively driven electrons with high enough longitudinal momentum such that they propagate away from the snowplow towards the vacuum-plasma boundary. The electrons reaching the vacuum-plasma interface create the slowly expanding sheath normal to the interface which then accelerates the heavy background-ions to kinetic energies up to 25MeV. It is seen that the snowplow electrons that are at oblique angles to the laser propagation gain higher momentum compared to the snowplow electrons on axis. This process is similar to ponderomotive swelling at oblique incidence\cite{ponderomotive-1999}. Because of the snowplow electrons obliquely propagating away from the snowplow, the transverse extent of the TNSA field is much bigger than the laser focal spot. In the simulations, even though the laser focal spot FWHM is only $7.6\mu m$, the sheath field is created over the whole $25.5\mu m$ transverse space which we simulate, as seen in Fig.\ref{fig:2D-TNSA-shock}[a].  The RITA accelerated protons get further accelerated by the TNSA field when they reach the vacuum-plasma interface. From Fig.\ref{fig:2D-TNSA-shock}[b] it is seen that the RITA proton bunch is accelerated to a higher energy by the TNSA field compared to its energy in Fig.\ref{fig:2D-proton-RITA-Snowplow}[a] ahead of the vacuum-plasma interface. Also, it is seen in Fig.\ref{fig:2D-TNSA-shock}[c] that the RITA proton bunch which has a smaller transverse size collides and accelerates beyond the TNSA proton bunch. The energy spectrum of the protons at this later time in Fig.\ref{fig:RITA-point1nc-energy-spectrum} shows the energy gained by the RITA accelerated proton bunch due to the TNSA fields.

\begin{figure}
	\begin{center}
   \includegraphics[width=3.6in]{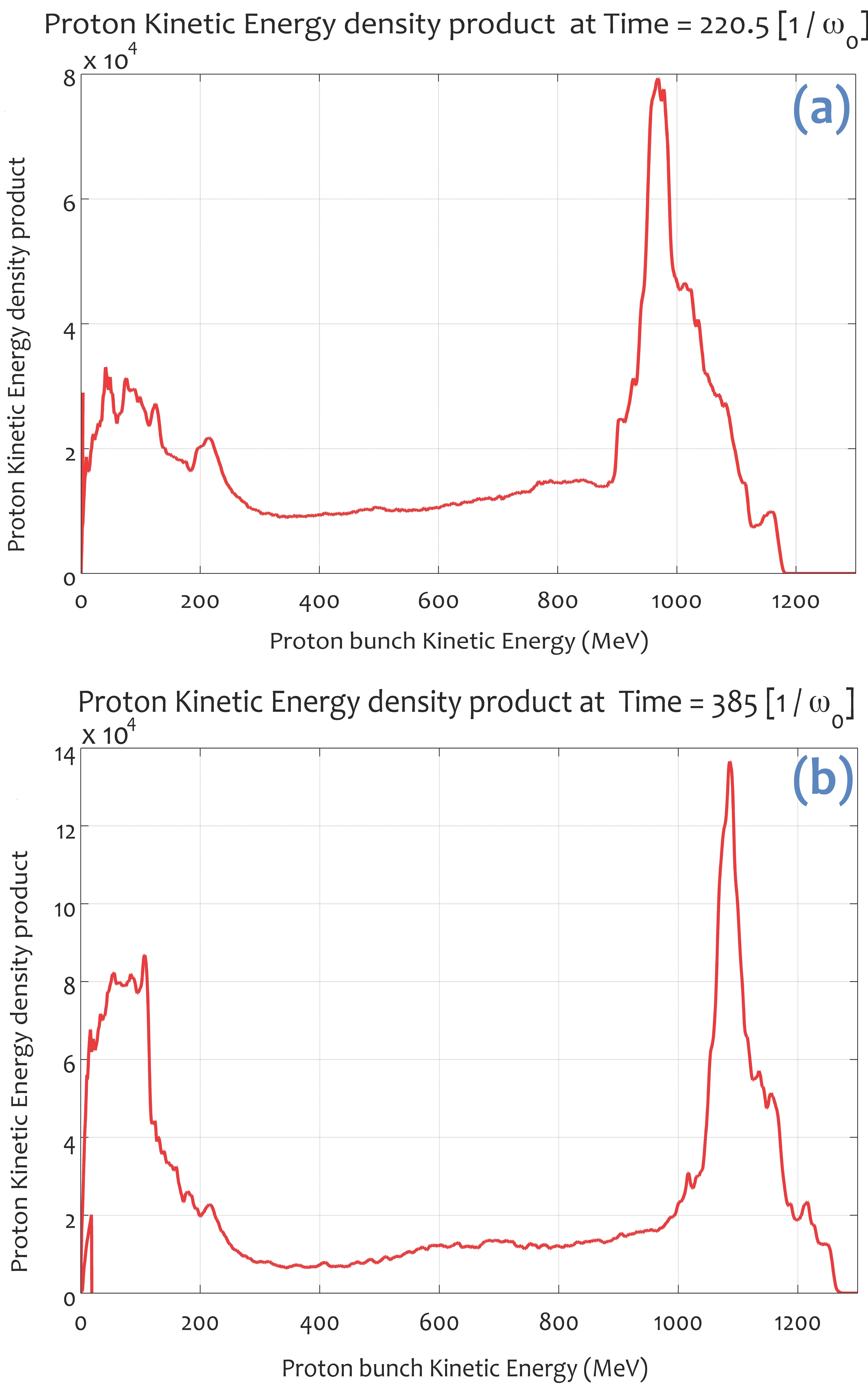}
	\end{center}
\caption{{\scriptsize {\bf Kinetic energy density product of the RITA accelerated proton bunch with $0.1\times n^{crit}_{cold}$.} {\bf a,} proton kinetic energy density at $t=220.5\frac{1}{\omega_0}=94fs$ for the laser-plasma parameters in Fig.\ref{fig:2D-snowplow-electron-density-evolution}. The peak energy is around 0.95 GeV with energy spectrum FWHM of around $10.7\%$. {\bf b,} proton kinetic energy density of the RITA accelerated protons after the second kick due to TNSA acceleration at $t=385\frac{1}{\omega_0}=164fs$. The RITA proton bunch has gained more than 100MeV or about $10\%$ of its initial energy due to its interaction with the TNSA field. It can be observed that the exponential energy distributed TNSA proton bunch is developing in the low-energy region of the spectrum. These spectrum graphs in (a) and (b) on a linear scale are the spatial integral of graphs in Fig.\ref{fig:2D-snowplow-electron-density-evolution}[a] and Fig.\ref{fig:2D-TNSA-shock}[b] respectively but for a higher proton trace density of $0.1\times n^{crit}_{cold}$ instead of  $0.01\times n^{crit}_{cold}$.}}
\label{fig:RITA-point1nc-energy-spectrum}
\end{figure}

\begin{figure*}[ht]
	\begin{center}
   \includegraphics[width=6in]{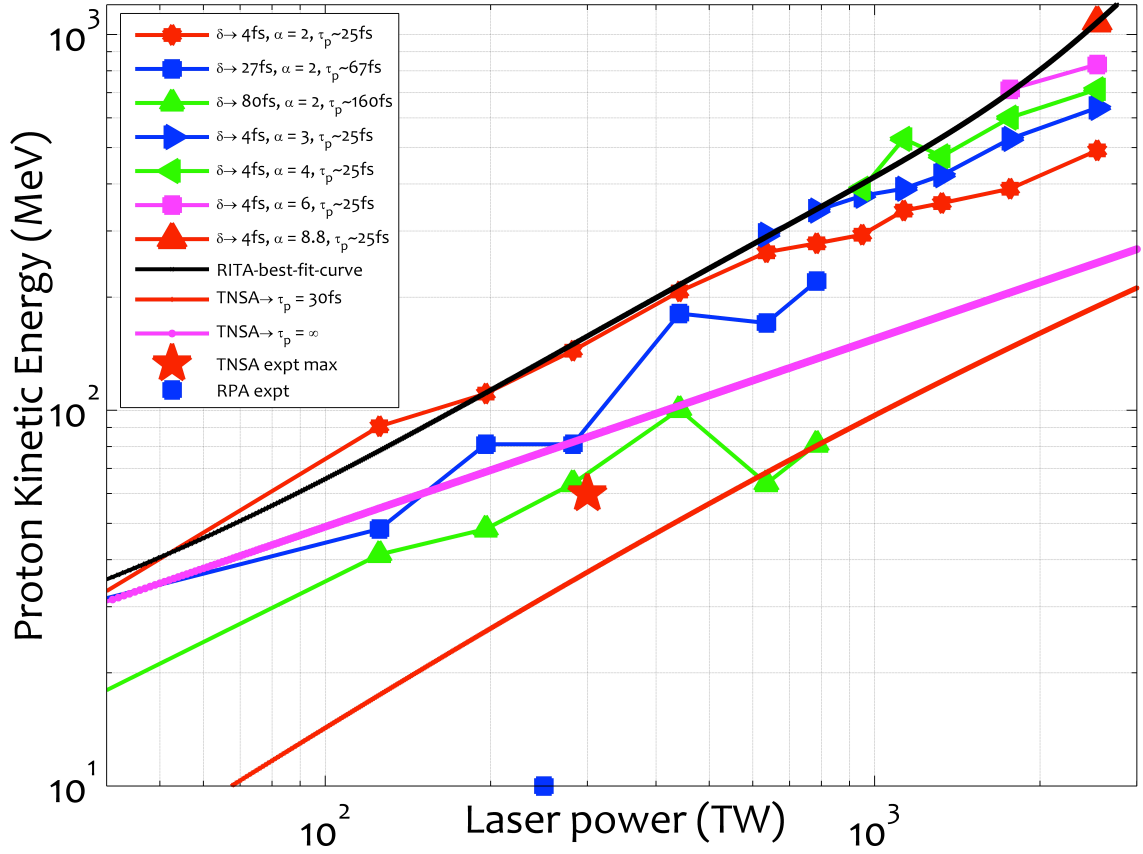}
	\end{center}
\caption{{\scriptsize {\bf Peak proton-bunch energy vs. Peak laser power.} A compilation of peak proton bunch energy obtained with 2-D simulations of the RITA scheme by varying the laser-plasma parameters, plasma density gradient scale-length $\alpha$ (in $\frac{c}{\omega_0}$), laser pulse rise-time scale-length $\delta$ and laser pulse $a_0$ (converted to laser peak power because the focal spot size $r_0$ is fixed across the compiled simulations) with fixed FWHM focal spot-radius, $r_0=3.8\mu m$. The top solid black line is a fit to the RITA 2-D energies for circularly polarized laser with rise-time $\frac{\delta}{c}\simeq 4fs$ and pulse length, $\tau_p=25fs$. The next two solid lines are for TNSA with $\tau_p=\infty$ and $\tau_p=30fs$ from \cite{tnsa-2006}\cite{scaling-laws-2010-ultrashort}. Experimentally obtained maximum TNSA \cite{tnsa-expt-max} and RPA \cite{rpa-mono-solid} proton energy are shown. }}
\label{fig:2D-snowplow-energy-vs-power-data}
\end{figure*}

To evaluate the validity of the RITA model in 2-D we summarize the laser-plasma parameter scaling law results in Fig.\ref{fig:2D-snowplow-energy-vs-power-data}. Here we compare the results (quasi-monoenergetic proton beams of different energies) of a number of 2-D simulations for varying laser-plasma parameters with fixed laser focal spot-size. It shows the predicted RITA scaling law that the proton beam energy, for a fixed intensity, increases directly with plasma density gradient scale-length ($\alpha$) and inversely with laser pulse rise-time ($\delta$). In Fig.\ref{fig:2D-snowplow-energy-vs-power-data} the RITA bunch energy, $E_p$, is plotted versus peak laser power, $P_0$, and is compared to TNSA (maximum cut-off energy of the exponential spectrum)\cite{tnsa-2006}\cite{scaling-laws-2010-ultrashort}\cite{tnsa-expt-max}. 2-D simulations for various intensities show that the proton beam energy scales linearly with laser power, $E_p \propto P_0$. We see that the RITA scaling is well above the TNSA maximum energy scaling, $E_p \propto \sqrt{P_0}$. The longer the TNSA field accelerates the ions, the higher the ion energies obtained\cite{tnsa-2006}, where the acceleration time $t_{acc}$ is directly proportional to laser pulse length $E_p \propto t_{acc} \propto \tau_p$. Also, the TNSA field amplitude depends upon the hot electron temperature, $T_e$ which is directly proportional to the laser pulse length, $\tau_p$. Hence the difference is greatest for short pulse lengths, since TNSA gives lower energy for shorter pulses while RITA is the opposite. Similarly in the RPA schemes\cite{rpa-2004} the ion energy scales directly with laser pulse-length $E_p \propto \tau_p$. In CESA scheme the shock Mach number scales directly with the shock electron temperature\cite{cesa-2004}. The shock electron temperature is shown to scale directly with the laser pulse length, thereby $E_p \propto \tau_p$. Hence RITA exhibits a unique inverse scaling of the accelerated ion energy with the laser pulse length (when $a_{plasma}>a_{th}$), $E_p \propto \frac{1}{\tau_p}$.

\section*{Acknowledgment}
The authors would like to acknowledge useful discussions with R. A. Bingham, P. K. Kaw,  C. J. Joshi, M. Tzoufras and S. C. Wilks. Work supported by the National Science Foundation under NSF-PHY-0936278, NSF-PHY-0936266 and NSF-PHY-0903039; the US Department of Energy under DEFC02-07ER41500, DE-FG02-92ER40727 and DE-FG52-09NA29552.


} 

\end{document}